\documentclass[a4paper]{article}
\usepackage[switch,pagewise]{lineno}
\usepackage[switch]{lineno}
\usepackage{icrc2013}
\usepackage{amssymb}
\usepackage{amsmath}
\usepackage{amsbsy}
\usepackage{grffile}
\usepackage[english]{babel}

\title{Coherent radio emission from the cosmic ray air shower sudden death}

\shorttitle{Emission from air shower sudden death}

\authors{Beno\^{\i}t Revenu and Vincent Marin}
\afiliations{SUBATECH, 4 rue Alfred Kastler, BP20722, 44307 Nantes, CEDEX 03, Universit\'e de Nantes, \'Ecole des Mines de Nantes, CNRS/IN2P3, France}

\email{revenu@in2p3.fr}

\abstract{
We describe the characteristics of a new radio signal, generated by the secondary electrons and positrons of the shower front when they reach the ground.
The very fast deceleration of these particles induces the coherent emission of an
electric field at frequencies smaller than 20~MHz. We show, using simulations with the code SELFAS, that this sudden death signal should be detectable with a simple dedicated antenna and could provide many informations on the shower, in particular the nature of the primary cosmic ray.
We also show that this signal permits to estimate the atmospheric depth
of maximum of electric field emission $X_\text{max}^\text{prod}$, which occurs well before the atmospheric depth corresponding to the maximum number of secondary particles in the shower ($X_\text{max}$). Observation of this signal should be considered for the design of future radio experiments.
}
\keywords{cosmic rays, extensive air showers, low frequency signal, sudden death}

\begin{document}


\maketitle


\section{Introduction}
The high energy cosmic rays (above $10^{16}$~eV) are studied through the extensive air showers (EAS) they generate when entering the atmosphere. The story of the shower starts at the first interaction point corresponding to the atmospheric depth $X_1$ (in g~cm$^{-2}$), then the shower develops in the atmosphere until the secondary particles in the shower front reach the ground level. It is possible to get informations on the primary cosmic ray during the development phase of the shower and also when the shower hits the ground, at an atmospheric depth $X_\text{ground}$.

{\bf {Development in the air}}

The shower development can be directly seen during dark moonless nights with the detection of the fluorescence light emitted by the nitrogen molecules excited by the shower. This signal allows to measure the longitudinal profile which provides in particular a calorimetric estimation of the primary energy and strong constraints on the nature of the primary cosmic ray. In addition to this optical signal, the secondary particles also emit an electric field that is detected by dedicated antennas. This electric field is at first order due to the geomagnetic mechanism through the systematic opposite drift of the electrons and positrons in the geomagnetic field~\cite{allan1971}. At second order, the relative excess of the electrons with respect to the positrons (the Askaryan effect in the air~\cite{ask1962}) is responsible for an additional contribution to the global electric field. These two mechanisms are studied in details in the AERA~\cite{huege_aera_icrc2013} experiment for instance. The CHerenkov radiation plays an important role close to the shower axis.

{\bf {At the ground level}}

When the shower front hits the ground, the secondary particles can induce a signal, for instance in water Cherenkov detectors (like in the Pierre Auger Observatory~\cite{Allekotte:2007sf}) or plastic sincitllators (like in the Telescope Array experiment~\cite{AbuZayyad201287}). This allows to estimate the ground particles densities and consequently, the shower core position and primary energy. In the radio domain, there should exist a signal associated to the end of the shower at the ground level. In this paper, we study this air shower sudden death mechanism and we argue that it should be coherent up to frequencies of the order of 20~MHz. In particular, the excess of electrons with respect to the positrons is the main source of this signal.

{\bf Radio detection of air showers at low frequencies}

 Previous experiments reported the observation of radio pulses undoubtly associated with air showers at low frequencies ($\leqslant 20$~MHz) but no satisfactory underlying mechanism has been proposed. The first detection of air shower in this frequency range has been obtained at a frequency of 2~MHz~\cite{1970Natur.225..253A}.

At 3.6~MHz~\cite{1970ICRC....3..717P,houghprescottclay1971}, it has been measured that the signal strength was a decade higher than the measurements in 20-60~MHz~\cite{allanclay1970} and three times smaller than at 2~MHz~\cite{stubbs1971}, where the electric field strength is estimated to $500~\mu$V~m$^{-1}$~MHz$^{-1}$ for showers at 10$^{17}$~eV, assuming a linear dependence between the electric field and the primary energy.
At 6~MHz~\cite{felgatestubbs1972}, the electric field has been recorded in both east-west (EW) and north-south (NS) polarizations and the authors concluded that the electric field was not entirely due to the geomagnetic mechanism. There was also a clear evidence for an increasing electric field strength with decreasing frequencies.

Using measurements at 22, 6, 2 and 0.1~MHz, it is argued in~\cite{claycrouchicrc1973,gregory1973} that a mechanism different from the geomagnetic one should operate, in particular at low frequencies.
It has been proposed, following~\cite{PhysRev.108.155,charman1968b} that the drift in the geoelectric field of the low energy electrons created by the ionization of the air after the passage of the shower, could be the main source of the measured electric field. This model does not produce electric fields higher than those produced through the geomagnetic mechanism~\cite{1978AuJPh..31..439S}.
Two other low-frequency emission mechanisms were proposed in~\cite{allan1972} but again, the upper limits calculated were much lower than the measured values.
In the Akeno experiment, radio signal in correlation with showers was demonstrated at frequencies in the range 26-300~kHz~\cite{1987ICRC....6..125N}. The radio pulses are monopolar and their amplitudes decrease as $1/d$ with the distance to the shower core. Contrarily to other experiments, the signal did not appear to be correlated with the geoelectric field amplitude. Distant showers, with core distance up to 2.5~km, have been observed with an electric field amplitude of $40~\mu$V/m. Various mechanisms are discussed and the most probable one is based on the electric field emission when the air shower electrons hit the ground, as discussed in~\cite{1985ICRC....7..308N,1985ICRC....7..268S,1983ICRC...11..428K}. Bipolar and monopolar pulses in the LF-MF band were also detected~\cite{1993ICRC....4..262K} in the AGASA experiment.

The EAS-RADIO experiment (installed at the EASTOP site) made measurements in the bands 350-500~kHz and 1.8-5~MHz~\cite{castagnoli1991}. The rising of the electric field strength with decreasing frequency is confirmed and field strengths of some hundreds of $\mu$V~m$^{-1}$~MHz$^{-1}$ have been measured at 470~kHz and 2.6~MHz.

The Gauhati group reported several measurements at 2 and 9~MHz~\cite{1993NCimC..16...17B}. They compared the results with the predicted values using a model based on the transition radiation mechanism from the excess negative charge of showers when hitting the ground. Their conclusion is that the transition radiation mechanism cannot explain all results and that another mechanism should exist.

The past results agree in a sense that the electric field associated to air showers has a (much?) higher amplitude at low frequencies (below 10~MHz) than in the range 20-80~MHz where the geomagnetic mechanism is known to be dominant. Other mechanisms have been tested, for instance: interaction between the ionization electrons in the air with the geoelectric field, transition radiation from the electrons in the shower front when reaching the ground, transverse and longitudinal emission assuming a full coherence but none of them can explain all measurements. The experimental results are not always consistent; this is mainly due, according to the authors, to the atmospheric conditions that were not properly recorded and can strongly affect the measured amplitudes.

{\bf Coherent deceleration of the shower front}

The new mechanism proposed in this paper is the coherent emission of the secondary electrons in excess in the shower front when reaching the ground level. These particles suffer a sudden deceleration when they disappear below the ground level; we will call this mechanism the sudden death (SD) mechanism, which creates the sudden death pulse (SDP).
At the macroscopic level, when the shower front hits the ground, the macroscopic charge density and current $\rho(\mathbf{r},t),~\mathbf{J}(\mathbf{r},t)$ vary very quickly. The resulting electric field depends on the time derivative of these quantities following the Jefimenko's equation~(see for instance~\cite{jackson}), and is consequently expected to reach high amplitudes.
At the microscopic level, this signal can be understood as coherent Bremsstrahlung, where the coherence is possible for frequencies smaller than 20~MHz, corresponding to wavelengths of the order of the size of the shower front.
We do not consider here the transition radiation emitted by the passage of the secondary charged particles through the boundary between air and ground but this mechanism will be taken into account for a more general treatment.

\section{Sudden death radio emission}

We use the code SELFAS~\cite{selfas2011}, where each secondary electron and positron of the shower front is considered as a moving source. The total electric field emitted by the complete shower at any observation point $\mathbf{r}$, is obtained after superposition of all individual contributions. In the Coulomb gauge, the total field detected by an observer located at $\mathbf{r}$ and at time $t$ is given by~\cite{selfas2011}:
\begin{equation}
\mathbf{E}_{tot}(\mathbf{r},t)=\frac{1}{4\pi\varepsilon_0 c}\frac{\partial}{\partial t}\sum _{i=1} ^{N_t}q_i(t_{\text{ret}})
\left[\frac{{\boldsymbol\beta}_{i}-(\mathbf{n}_i.\boldsymbol{\beta}_{i})\mathbf{n}_i}{R_i\,(1-\eta\boldsymbol{\beta}_i.\mathbf{n}_i)}\right]_{\text{ret}}
\label{SumField}
\end{equation}
where $\eta$ is the air refractive index, $\mathbf{n}_i$ and $R_i$ are the line of sight and the distance between the observer and the particle $i$, $\boldsymbol{\beta}_i$ the velocity of this particle and $q_i$ its electric charge. The summation is done over the total number $N_t$ of particles that emitted an electric field detected by the observer at time~$t$.
All these quantities are evaluated at the retarded time $t_{\text{ret}}$, related to the observer's time $t$ through
$t=t_{\text{ret}}+\eta \,R_i(t_{\text{ret}})/c$.
The Earth's magnetic field induces a systematic opposite drift of electrons and positrons during the shower development; this generates a residual current
perpendicular to the shower axis. The variation of this current with the number of particles creates the main electric field contribution, at frequencies above $\sim30$~MHz. This is known as the geomagnetic mechanism and the charge excess variation gives a secondary contribution to the total electric field, as discussed previously. The description adopted in SELFAS takes into account both mechanisms. Experimental evidence for the charge excess contribution can be found in~\cite{Schoorlemmer2012S134,marinicrc2011,huege_aera_icrc2013}.
The resulting electric field generated during the shower development in the air creates the principal pulse (PP), as already observed since years by several experiments such as CODALEMA~\cite{Ardouin:2005xm,coda2009},  LOPES~\cite{lopes1,Huege2012S72} or AERA~\cite{Abreu:2011ph} for the most recent ones.

\begin{figure}[!ht]
\begin{center}
\includegraphics[scale=0.3]{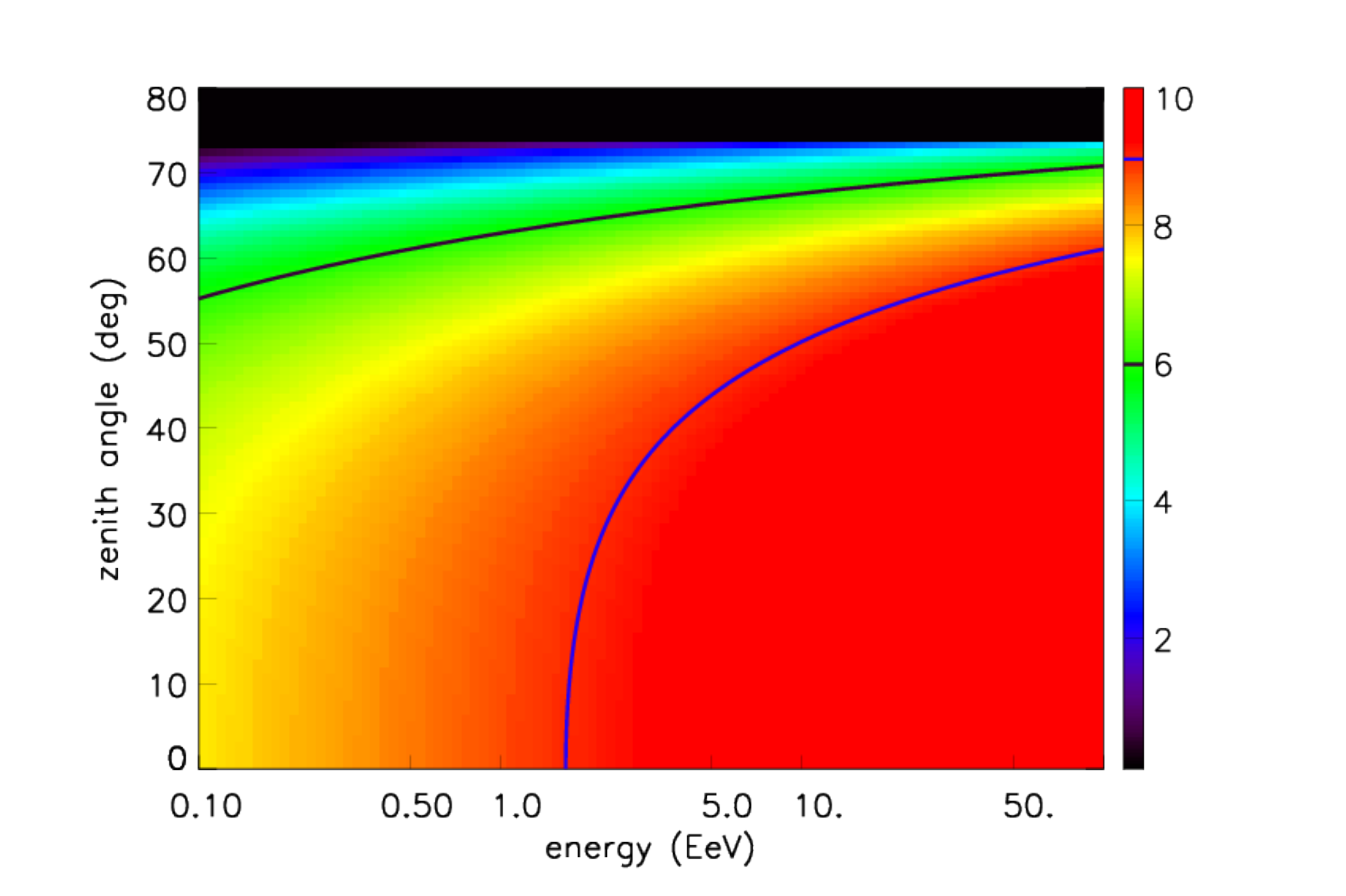}
\end{center}
\caption{Number of particles reaching the ground $N(X_{\text{grd}})$ as a function of zenith angle and energy, computed for an altitude of 1400~m assuming proton showers. The color scale indicates the $\log_\text{10}(N(X_{\text{grd}}))$. The contours for $10^6$ and $10^9$ ground particles are overplotted.}
\label{PartGround}
\end{figure}

The characteristics of the electric field created by the shower front sudden death depends on the ground distribution of the shower electrons and positrons but also on the altitude of the observation site. The number of secondary particles reaching the ground $N(X_{\text{grd}})$ can vary strongly according to
the energy of the primary cosmic ray inducing the cascade in the atmosphere, the air shower arrival direction (zenith angle),
 the first interaction length $X_1$ of the primary cosmic ray and the nature of the primary cosmic ray.
Fig.~\ref{PartGround} shows the number of particles reaching the ground as a function of arrival direction, for showers induced by protons, assuming an altitude of 1400~m corresponding to the Auger site. We used the Greisen-Iljina-Linsley parameterization~\cite{GIL,Greisen} of the longitudinal profile.
The maximum number of particles $N_{\text{max}}$ in the shower is obtained at the depth of maximum development $X_{\text{max}}$. We see that for a primary energy above $10^{18}$ eV (1~EeV) and for arrival directions $\theta\leqslant50^\circ$, $N(X_{\text{grd}})$ is not negligible with respect to $N_{\text{max}}$ ($N(X_{\text{grd}})/N_{\text{max}}\geqslant10\%$).
A large majority of secondary electrons and positrons (at least 97\%) for $X\geqslant X_{\text{max}}$, have an energy smaller than 1~GeV (see~\cite{lafebre2009}). Consequently, when the shower front hits the ground, the secondary electrons and positrons are stopped on a distance smaller than 1~m at 1~GeV in a silicon medium~\cite{Nist}, similiar to the ground at the Auger site.
The SD mechanism operates in a region close to the shower core: an observer will therefore detect the signal at a time~$\sim d/c$ after the impact on the ground, where $d$ is the distance to the shower core. The electric field emitted during the development of the shower in the atmosphere creates the PP as previously discussed, which is detected before this sudden death pulse. The time interval between the PP and the SDP depends on the geometry (the air shower arrival direction and the antenna position with respect to the ground shower core) but also on the instant of the maximum of emission during the air shower development. 

\section{Description of the pulse}

\begin{figure}[!ht]
\begin{center}
\includegraphics[width=8cm]{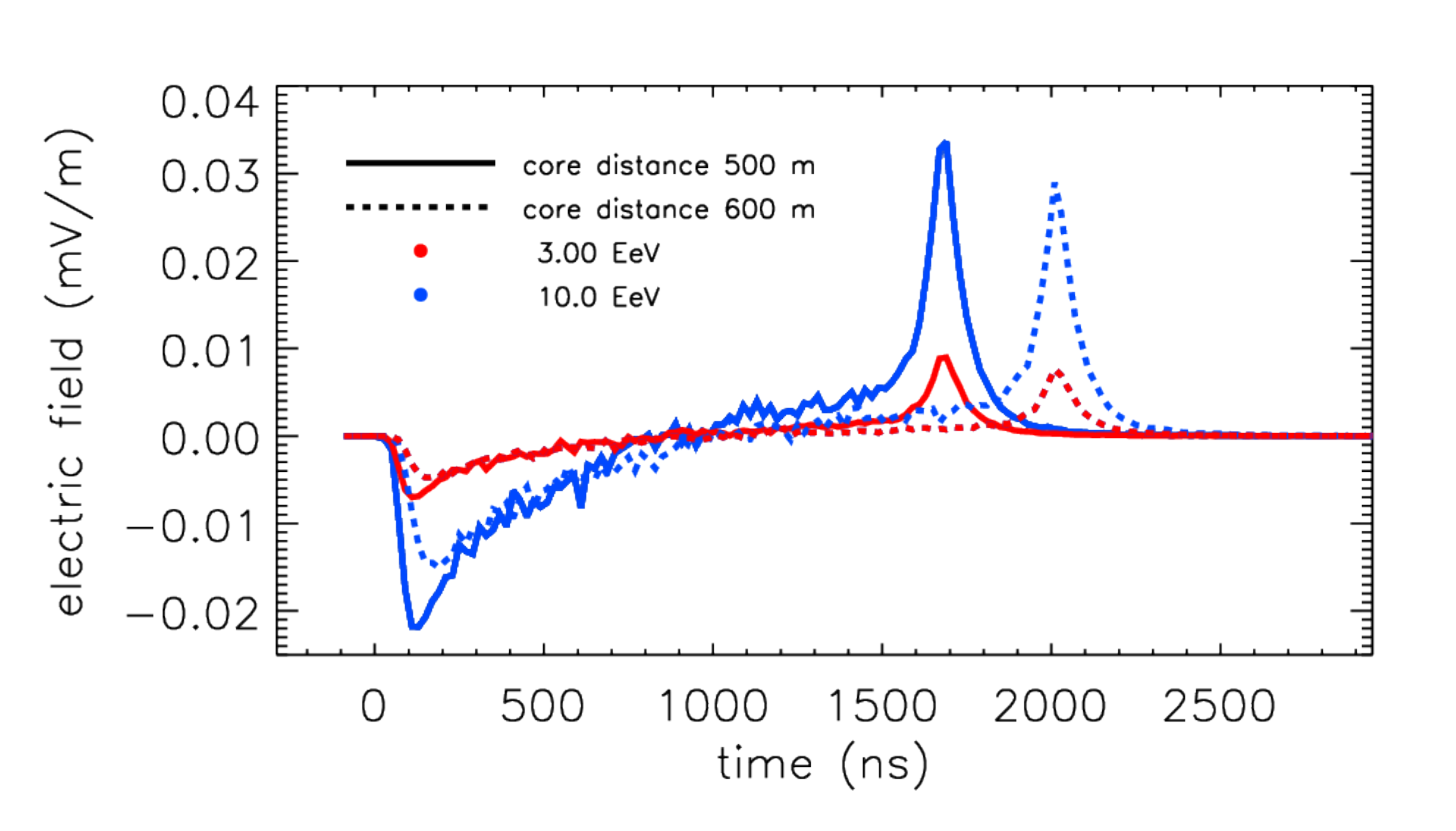}
\end{center}
\caption{V electric field vs time for two antennas located at 500~m and 600~m at the east of the shower core, for a vertical showers at 3 and 10~EeV. The PP are clearly visible at the beginning of the traces, around 100~ns and the SDP pulses are located around 1600~ns and 2000~ns. The SDP pulses don't exist in the EW and NS polarization, in agreement with the predicted polarization pattern of Eq.~\ref{SumField}.}
\label{pulseEWbrasN}
\end{figure}

We ran simulations of proton-induced vertical and inclined showers falling at the origin of the coordinate system for energies $0.1, 1, 3, 10, 100$~EeV with the code SELFAS for 48 antennas located between 100~m and 800~m by steps of 50~m at the geographic east, west and north of the shower core and one antenna located at the origin. The ground level is set at 1400~m and we consider the geomagnetic field measured at the Pierre Auger Observatory.
The first interaction point $X_1$ at a given energy is extracted from the QGSJET~\cite{Ostapchenko2006143} data, and is taken as a function of energy as $52.8, 48.6, 46.8, 44.9, 41.5$~g~cm$^{-2}$, respectively. The electric field received by each antenna as a function of time is calculated in the EW, north-south (NS) and vertical (V) polarizations.
The observer's origin of time $t=0$ corresponds to the time when the shower axis hits the ground at the core position~$(0,0,1400)$.
Fig.~\ref{pulseEWbrasN} presents the V electric field as a function of time at 500~m and 600~m at the east of the shower core. We obtain similar figures for the V and NS polarizations.
The PP due to the various mechanisms generating the electric field in the shower during the development in the atmosphere is clearly visible at the beginning of the trace, with a maximum amplitude occuring at a time close to -500~ns in this example.
The SDP appears at roughly 1600~ns$=500~\text{m}/c$ and 2000~ns $=600~\text{m}/c$ for the 2 antennas considered here.
We checked that the contribution to the SDP is mainly due to the excess of negative charge in the shower.

From the simulation, the time interval between the time origin and the time
when the SDP reaches its maximum value is simply given by $\delta t=d/c$, in agreement with the hypothesis that the SDP is due to the disappearance of secondary particles when hitting the ground. This result is also verified for all antennas used in this simulation. The SDP is monopolar because it is computed as the time derivative of a decreasing current density to 0 ($\beta=0$ when the particles are below the ground level). The symetric shape of the pulse can be explained by the symetric shape of the lateral distribution function.

\begin{figure}[!ht]
\begin{center}
\includegraphics[width=8cm]{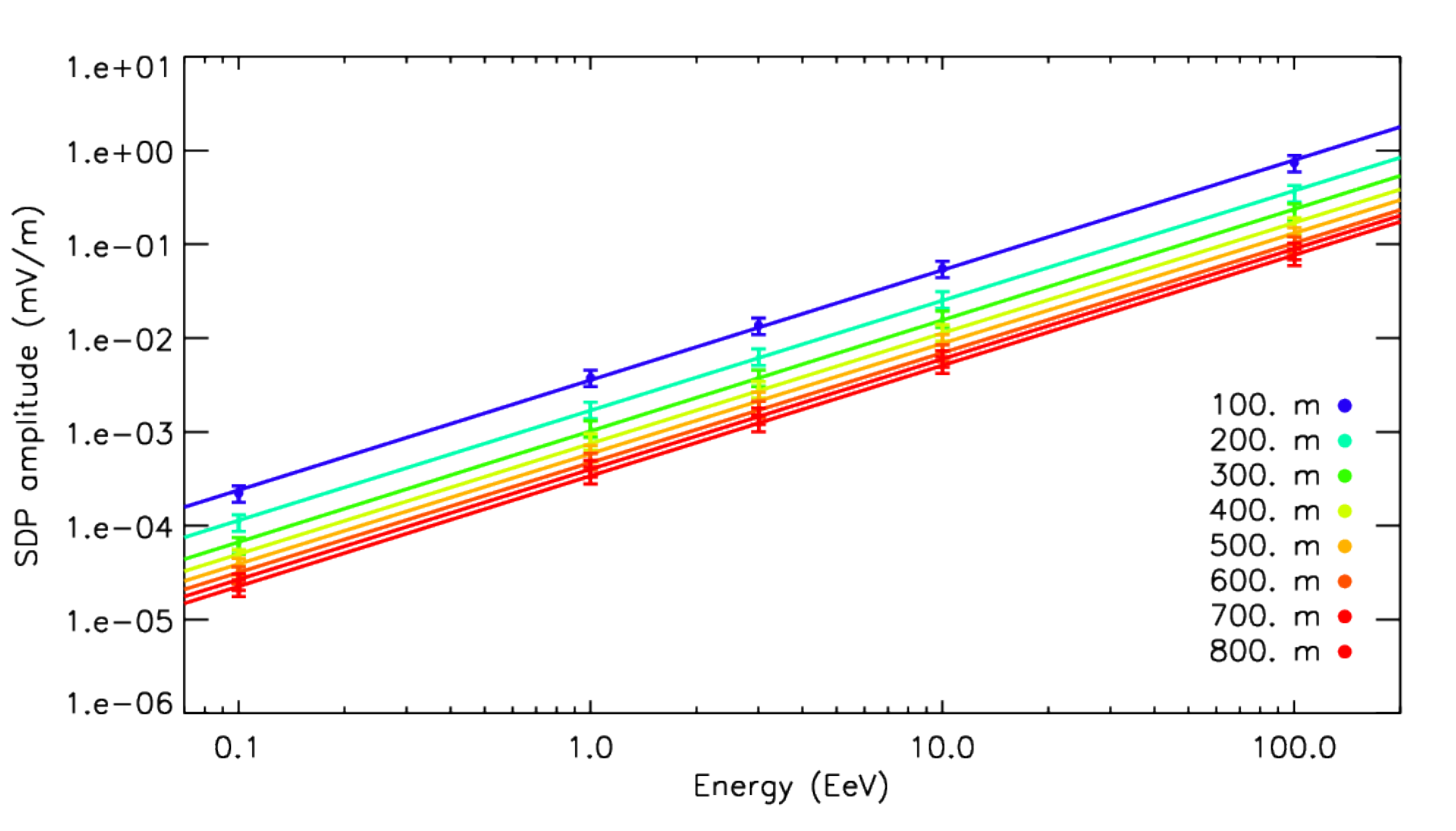}
\end{center}
\caption{SDP amplitude in the V polarization as a function of energy for different core distances. The antennas are located at the east of the shower core. The SDP amplitude varies almost linearly with the energy. This case corresponds to an inclined shower with $\theta=30^\circ$ and $\phi=45^\circ$.}
\label{calibinclined_energy}
\end{figure}

We obtained that the amplitude scales linearly with the primary energy and as $1/d$ where $d$ is the core distance. The same result holds for vertical showers. The reference amplitude at 1~EeV for a core distance of 100~m is of the order of 5~$\mu$V/m for a zenith angle of 30$^\circ$.
Fig.~\ref{calibinclined_energy} presents the variation of the SDP amplitude for an inclined shower with $\theta=30^\circ,~\phi=45^\circ$, as a function of energy for different core distances. The amplitude varies linearly with the primary energy and scales as $1/d$ where $d$ is the core distance. In comparison, the PP decreases exponentially with axis distance.

\subsection{Frequency domain}

\begin{figure}[!ht]
\begin{center}
\includegraphics[width=8cm]{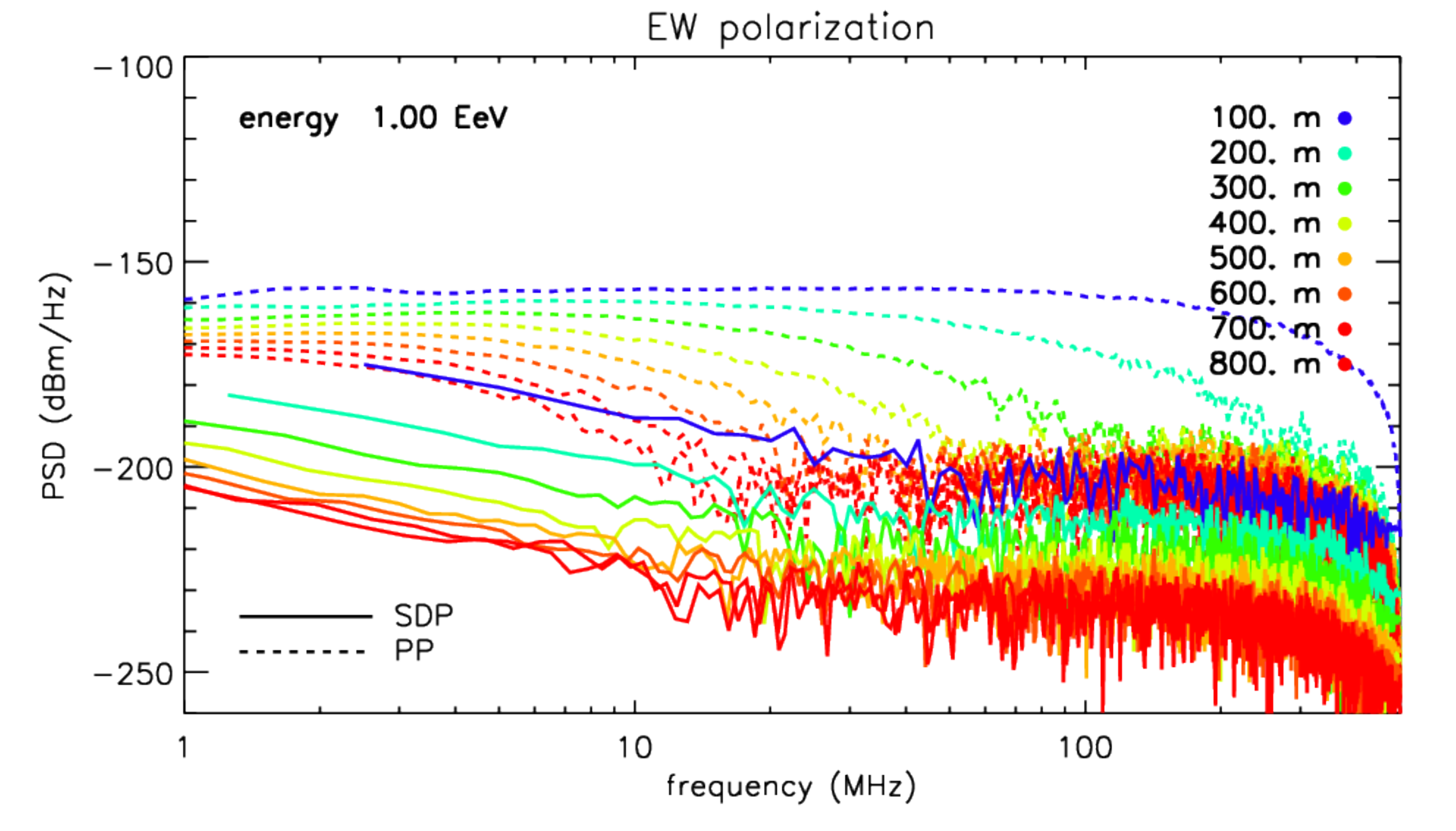}
\end{center}
\caption{\footnotesize{PSD of the PP and SDP in the EW polarization for different core distances located at the east of the shower axis. The simulated shower has an energy of 1~EeV with $\theta=30^\circ,~\phi=45^\circ$.}}
\label{sdpsd}
\end{figure}

Contrarily to the time structure of the PP, that of the SDP does not depend on the primary energy nor on the distance to the shower core as can be seen in Figs.~\ref{pulseEWbrasN}. Only the amplitude is affected by these parameters. Therefore, we expect the shape of the power spectral density (PSD) to be similar at all core distances and all energies as a function of frequency, the only difference being the normalization of the PSD. Fig.~\ref{sdpsd} presents the PSD of the PP for a primary energy of 1~EeV and core distances varying from 100~m to 800~m. We observe, as already predicted by several simulation codes, that the PP is coherent up to a certain frequency depending on the axis distance of the observer.
This is not the case for the SDP since we show that it is coherent up to a frequency of $\sim20$~MHz independently of  the core distance.

\section{Discussion and conclusion}

\begin{figure}[!ht]
\begin{center}
\includegraphics[width=9cm]{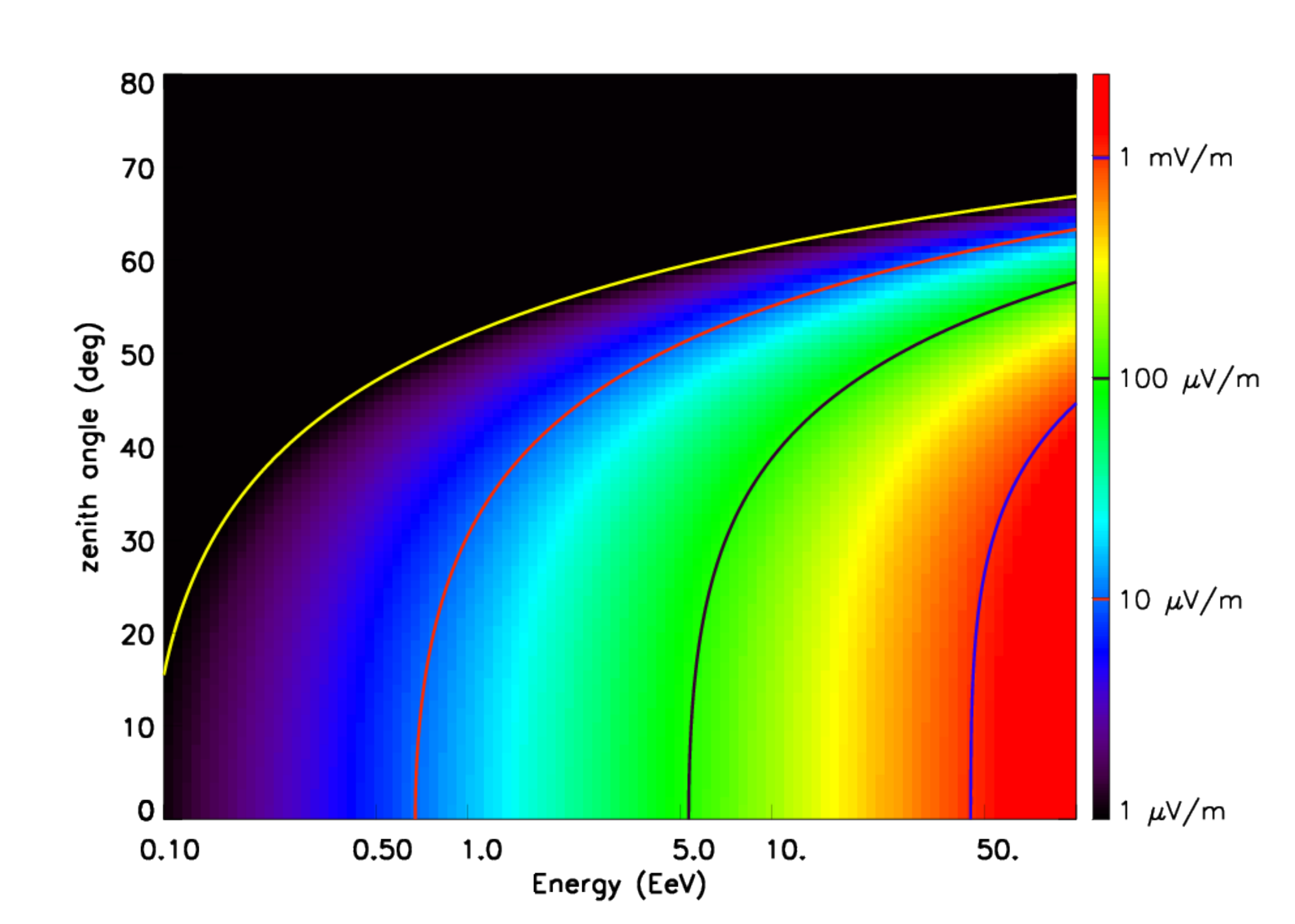}
\end{center}
\caption{\footnotesize{Expected amplitude of the SDP at 100~m at the east of the shower core, in the EW polarization, as a function of primary energy and zenith angle, computed for the site of the Pierre Auger Observatory.}}
\label{ampEt}
\end{figure}

We have characterized the electric field produced by the secondary $e^\pm$ when they reach the ground. The amplitude of this signal in the horizontal polarization, is of the order of $15~\mu$V/m at 100~m of the shower core for a vertical shower,
and a primary energy of 1~EeV. The amplitude scales linearly with the energy and decreases as $1/d$, where $d$ is the distance between the shower core and the observer. The lateral distribution function for the SDP is less steep than the lateral distribution function for the PP. This could explain why previous experiments reported air shower detection at higher distances when working at low frequencies. The SDP amplitude depends on the distance to the shower core, contrarily to the PP due to the secondary $e^\pm$ during the shower development, which strongly depends on the distance to the shower axis. The polarization is oriented along the vector ${\boldsymbol\beta}-(\mathbf{n}.\boldsymbol{\beta})\mathbf{n}$, where $\mathbf{n}$ is the normalized vector between the shower core and the observer position and $\boldsymbol\beta$ can be taken as the shower axis direction. The predicted amplitude depends on the zenith angle and primary energy of the shower. Fig.~\ref{ampEt} presents the expected SDP amplitude at 100~m at the east of the shower core, in the EW polarization, as a function of zenith angle and energy. We assumed proton-initiated showers with the first interaction point $X_1$ set to the average value extracted from QGSJET data as a function of primary energy.
If we are able to detect a signal at these frequencies, then it would be quite simple to trigger on actual cosmic rays because any trace having two pulses separated by some $\mu$s could be an excellent candidate. This would be a very specific signature of cosmic rays and would help a lot in discarding background events. The SDP is expected to arrive at the observer location at a delayed time $d/c$ with respect to the core time. The core is therefore located on a circle centered on the observer position and of radius $d/c$. Using the information from several detectors, the shower core is at the intersection of the corresponding circles. The SDP amplitude is proportionnal to the total number of secondary $e^\pm$ and also reflects their complete ground distribution, contrarily to particle detectors that sample at specific locations this ground distribution. It means that the SDP is specifically sensitive to the full electromagnetic ground component of the shower.
Experimentally, one should be interested in recording the electric field over a duration greater than some $\mu$s to be able to observe the SDP (up to $\sim3~\mu$s after the impact time at 1~km from the shower core). The antennas used should be also sensitive to horizontal directions and vertical polarization.

Because the SDP is generated by the end of the shower, its detection provides an absolute timing of the shower in the trace recorded by an antenna. For antennas not too close from the shower axis (because the effect of the air refractive index is important close to the shower axis), we expect a one-to-one correspondence
between the time in the antenna trace and the position of the source in the shower. This bijective relation can be established by simple geometrical considerations and it is therefore possible to determine the atmospheric depth $X_{\text{max}}^{\text{prod}}$ corresponding to the maximal production of electric field. Using simulations, we obtained that $X_{\text{max}}^{\text{prod}}$ is close to the atmospheric depth where the secondary particle production rate of the shower is maximum (ie the inflexion point of the longitudinal profile). The details will be explained in a forthcoming publication where the transition radiation will be included.

\end{document}